%% file: main.tex
\documentclass[conference]{IEEEtran}

\usepackage{amsmath,amsfonts}
\usepackage{esvect}
\usepackage{graphicx}
\usepackage{subfigure}
\usepackage{booktabs}
\usepackage{tabularx}
\usepackage{multirow}
\usepackage{balance}
\usepackage{graphicx}
\usepackage{multirow}
\usepackage{xcolor}
\usepackage{enumitem}
\usepackage{mathtools}
\usepackage{wrapfig}
\usepackage{ragged2e}
\usepackage[all]{nowidow}
\usepackage{stfloats}
\usepackage{anyfontsize}
\usepackage{color}
\usepackage{colortbl}
\usepackage{mathtools}
\usepackage{longtable}
\usepackage{lscape}
\usepackage[linesnumbered,ruled,vlined]{algorithm2e}
\usepackage{syntax}
\usepackage{pifont} 
\usepackage{url}
\usepackage{listings}
\usepackage{xcolor}

\newcommand{\comm}[1]{{#1}}

\newcommand{\sectopic}[1]{\vspace*{0.1em}\par\noindent{\textit{\bfseries #1}}}

\definecolor{customBlue}{RGB}{68, 114, 196}
\newcommand {\StepOne}{\color{customBlue}\ding{202}\color{black}}
\newcommand {\StepTwo}{\color{customBlue}\ding{203}\color{black}}
\newcommand {\StepThree}{\color{customBlue}\ding{204}\color{black}}
\newcommand {\StepFour}{\color{customBlue}\ding{205}\color{black}}
\newcommand {\StepFive}{\color{customBlue}\ding{206}\color{black}}

\definecolor{customGreen}{RGB}{83, 129, 53}
\newcommand {\StepOneAnswerGeneration}{\color{customGreen}\ding{202}\color{black}}
\newcommand {\StepTwoAnswerGeneration}{\color{customGreen}\ding{203}\color{black}}

\usepackage{listings}

\begin{document}

\title{\comm{Developing a Llama-Based Chatbot for CI/CD Question Answering: A Case Study at Ericsson}}

\author{
    \IEEEauthorblockN{%
    Daksh Chaudhary\IEEEauthorrefmark{1}\IEEEauthorrefmark{2}, 
    Sri Lakshmi Vadlamani\IEEEauthorrefmark{2},
    Dimple Thomas\IEEEauthorrefmark{2},
    Shiva Nejati\IEEEauthorrefmark{1}, 
    Mehrdad Sabetzadeh\IEEEauthorrefmark{1}
   }
    \IEEEauthorblockA{\IEEEauthorrefmark{1}University of Ottawa, 800 King Edward Avenue, Ottawa ON K1N 6N5, Canada}
    \IEEEauthorblockA{\IEEEauthorrefmark{2}Ericsson Canada, 349 Terry Fox Dr, Kanata, ON K2K 2V6
    }
	Email: \{dchau012, snejati,  m.sabetzadeh\}@uottawa.ca; \{sri.lakshmi.vadlamani, dimple.thomas\}@ericsson.com
}

\maketitle

\input{abstract}

\input{intro}
\input{context}
\input{background}
\input{chatbot}
\input{exp}
\input{lesson}
\input{conclusion}
{%
\appendix

\subsection{Parameters for Microsoft Teams Messages}\label{app:messages} As of this writing, Microsoft Teams retains the following 35 parameters (attributes) for each message sent:\\
{\footnotesize\tt%
id, 
etag, 
messageType, 
createdDateTime, 
lastModifiedDateTime, 
lastEditedDateTime, 
importance, 
locale, 
webUrl, 
attachments, 
mentions, 
reactions, 
from.user.@odata.type, 
userId, 
userDisplayName, 
userIdentityType, 
tenantId, 
contentType, 
content, 
channelIdentity.teamId, 
channelIdentity.channelId, 
subject, 
deletedDateTime, 
eventDetail.@odata.type, 
eventDetail.channelId, 
eventDetail.channelDescription, 
eventDetail.initiator.application, 
eventDetail.initiator.device, 
eventDetail.initiator.user.@odata.type, 
eventDetail.initiator.user.id, 
eventDetail.initiator.user.displayName, 
eventDetail.initiator.user.userIdentityType, 
eventDetail.channelDisplayName, 
eventDetail.visibleHistoryStartDateTime, 
eventDetail.members}.

\subsection{Parameters for Microsoft Teams Replies}\label{app:replies}

As of this writing, Microsoft Teams retains the following 43 parameters (attributes) for each reply to a message:\\{\footnotesize\tt%
id, 
replyToId, 
etag, 
messageType, 
createdDateTime, 
lastModifiedDateTime, 
lastEditedDateTime, 
deletedDateTime, 
subject, 
summary, 
chatId, 
importance, 
locale, 
webUrl, 
onBehalfOf, 
policyViolation, 
eventDetail, 
attachments, 
mentions, 
reactions, 
from.application, 
from.device, 
from.user.@odata.type, 
userId, 
userDisplayName, 
userIdentityType, 
tenantId, 
contentType, 
content, 
channelIdentity.teamId, 
channelIdentity.channelId, 
from, 
eventDetail.@odata.type, 
eventDetail.callId, 
eventDetail.callDuration, 
eventDetail.callEventType, 
eventDetail.callParticipants, 
eventDetail.initiator.application, 
eventDetail.initiator.device, 
eventDetail.initiator.user.@odata.type, 
eventDetail.initiator.user.id, 
eventDetail.initiator.user.displayName, 
eventDetail.initiator.user.userIdentityType}.

\section*{Acknowledgements}
We are grateful to the anonymous reviewers of the ICSME 2024 Industry Track for their insightful feedback. We further thank Dorian Gerdes for his valuable suggestions.}

\balance

\bibliographystyle{plain}
\bibliography{ref}

\end{document}

%% file: abstract.tex
\begin{abstract}
This paper presents our experience developing a Llama-based chatbot \comm{for question answering about} continuous integration and continuous delivery (CI/CD) at Ericsson, a multinational telecommunications company. Our chatbot is designed to handle the specificities of CI/CD \comm{documents} at Ericsson, employing a retrieval-augmented generation (RAG) model to enhance accuracy and relevance. Our empirical evaluation of the chatbot on industrial CI/CD-related questions indicates that an ensemble retriever, combining BM25 and embedding retrievers, yields the best performance. 
When evaluated against a ground truth of \comm{72 CI/CD questions and answers} at Ericsson, our most accurate   chatbot configuration provides fully correct answers for 61.11\% of the questions, partially correct answers for 26.39\%, and incorrect answers for 12.50\%. Through an error analysis of the partially correct and incorrect answers, we discuss the underlying causes of inaccuracies and provide insights for further refinement. We also reflect on lessons learned and suggest future directions for further improving our chatbot's accuracy.
\end{abstract}

%% file: intro.tex
\section{Introduction}
With advances in large language models (LLMs), the high-tech industry is increasingly looking into how chatbots can improve software engineering practices. 
There have been existing attempts to employ LLM-based chatbots in the domain of software engineering. Among others, Abdellatif et al.~\cite{Abdellatif2022, abdellatif2020challengesinchatbot} and Daniel \& Cabot~\cite{DANIEL2024103032} explore various facets of integrating chatbots into software engineering workflows, highlighting the potential for chatbots to streamline processes, enhance communication, and assist in tasks ranging from bug tracking and documentation to code generation and quality assurance.

This paper presents our experience developing a Llama-based chatbot \comm{for answering questions related to} continuous integration and continuous delivery (CI/CD) at Ericsson, a multinational telecommunications company. Ericsson employs agile development and DevOps across various projects, requiring many engineers to work efficiently with CI/CD processes.

CI/CD enables automated testing, integration, and deployment of code changes~\cite{humble2010continuous, rossel2017continuous, belmont2018hands}. CI/CD is intrinsically linked to software maintenance and evolution, ensuring that software remains functional and up-to-date as new features and fixes are continuously integrated into the codebase. Our chatbot is designed to handle the contextual factors specific to CI/CD \comm{documents} at Ericsson. This includes the evolving content of guidelines and team conversations about CI/CD.

To build an accurate chatbot, we opt for a retrieval-augmented generation (RAG) model~\cite{lewisrag}. RAG enhances chatbot capabilities by combining retrieval of relevant documents with the generative power of LLMs, thereby providing more accurate and relevant responses. RAG presents two main advantages over fine-tuning a model on domain-specific corpora~\cite{ezzini2023aibased}: First, fine-tuning requires a labeled dataset, which can be expensive to build. Second, a fine-tuned model is prone to outdated knowledge; this issue can be mitigated through a RAG model that continuously accesses and retrieves up-to-date information.

We evaluate our chatbot on industrial CI/CD questions. We experiment with alternative retriever models for instantiating a RAG pipeline over Llama 2 and empirically evaluate the accuracy of the resulting pipelines. Our evaluation indicates that an ensemble retriever, combining BM25~\cite{BM25} and embeddings~\cite{bengio2014representation,Mitra2018AnIT}, leads to the best overall outcome. Specifically, by using an ensemble retriever and evaluating the chatbot against a \comm{ground truth of 72 CI/CD questions and answers} at Ericsson, we obtain fully correct answers for 61.11\% of the questions, partially correct answers for 26.39\%, and incorrect answers for 12.50\%. Following this evaluation, we conduct an error analysis on partially correct and incorrect answers to identify the root causes of the inaccuracies.

\textbf{Novelty.} The novelty of our work lies in providing practical insights into the readiness of chatbot technologies for \comm{question answering} in a complex and specialized yet dynamic setting, where the content from which answers are derived is fluid and changes over time. To the best of our knowledge, we are the first to report on the construction and evaluation of a chatbot for \comm{answering CI/CD questions} in an industrial context.

\textbf{Significance.} Despite recent advances in generative LLMs that have made chatbot construction more accessible, the field remains marked by hype, generally lacks empirical analysis of accuracy, and does not sufficiently elaborate on technical considerations that could have a make-or-break effect on chatbot efficacy. Our work highlights our main technical choices for chatbot design, aiming to assist other researchers and practitioners facing similar challenges. Furthermore, we provide an empirically grounded examination of chatbot accuracy for a software-engineering problem in industry. This contributes to the development of a scientific body of knowledge that facilitates wider deployment of software-engineering chatbots.

\textbf{Structure.} The rest of this paper is structured as follows: Section~\ref{sec:context} motivates our work and presents our industry context.
Section~\ref{sec:background} provides background information. Section~\ref{sec:related} surveys related work. Section~\ref{sec:approach} describes the technical approach that underlies our chatbot. Section~\ref{sec:evaluation} reports on the evaluation of our chatbot. Section~\ref{sec:lessons} discusses lessons learned. Section~\ref{sec:conclusion} concludes the paper.

%% file: context.tex
\section{Industrial Context and Motivation}\label{sec:context}
Ericsson is a multinational company specializing in providing ICT services and equipment to telecommunications operators and enterprises worldwide. Our chatbot was developed within Ericsson's Cloud Radio Access Network (CloudRAN) unit. This unit focuses on online and virtualized central system observability and monitoring solutions across cloud-native RAN deployments, including software microservices on containers-as-a-service (CaaS) infrastructure. 

CloudRAN employs CI/CD to automate code integration, testing, and delivery. The CI/CD process at CloudRAN adheres to industry best practices and follows a standard workflow~\cite{cicdericsson}: All code is stored in a version-control system, with developers working on feature branches. When code is committed, a CI server triggers automated builds and tests to provide instantaneous feedback. Successful builds produce artifacts for deployment. During the CD process, code that passes CI tests is deployed to a staging environment for additional testing and manual review. CloudRAN has an approval process in place before deploying to production. Monitoring tools and centralized logging systems track application performance and detect issues. This iterative process ensures frequent, reliable code integration and delivery, reducing errors and downtime while enabling rapid feedback and improvements.

Our chatbot aims to improve CI/CD at CloudRAN by enabling software engineers, both within the unit and at client sites, to \comm{obtain answers to their CI/CD-related questions}. A few examples of queries that CloudRAN would like the chatbot to be able to respond to are: \emph{%
(1)~What are the steps to release a microservice?
(2)~How can I modify test targets during staging?
(3)~How can I migrate from [cluster~1] to [cluster~2]?
(4)~How do I add a test channel to a  Jenkins pool?}

We note that, both in the above queries and in the examples provided throughout the rest of the paper, we have altered the content from its original form to preserve confidentiality, while ensuring that the substance remains unchanged. Any redacted text is indicated by square brackets ([]).

By handling routine queries, such as our illustrative examples above, the chatbot offers the potential to free up expert engineering resources to address more complex issues. This reduces operational costs, speeds up issue resolution, and allows experts to focus on tasks that require specialized skills.

The dynamic nature of CloudRAN's operations, which includes reliance on constantly evolving internal documents and team communications, is an important contextual factor to consider in the design of the chatbot to ensure its longevity. To this end, we employ RAG to incorporate up-to-date information from team workspaces and messaging channels, providing answers based on the most recent knowledge.

%% file: background.tex
\section{Background}\label{sec:background}
Our chatbot falls under the umbrella of retrieval-augmented question-answering (QA) techniques. Retrieval-augmented QA involves integrating a retrieval mechanism to extract pertinent information from a given set of documents, thereby enhancing the accuracy and completeness of answers to queries. Retrieval augmentation has been explored for both extractive QA~\cite{ezzini2023aibased} and generative QA~\cite{Semnani2023WikiChat, liu2023retallm}, with significant accuracy improvements shown for both types of QA. For generative QA, which is the focus of our work, the associated prompt engineering is one of the most critical steps. OpenAI provides several general guidelines on how to build effective prompts~\cite{openaiprompt} for RAG tasks. We follow these guidelines for building retrieval augmentation into our chatbot.

A common approach for implementing retrieval augmentation is through a \emph{retriever and reader architecture}~\cite{ezzini2023aibased}. The retriever is responsible for efficiently selecting a subset of relevant documents from a larger corpus, acting as an initial filter to narrow the search space. Subsequently, the reader -- typically an LLM -- is tasked with comprehending and extracting/generating information based on the retrieved documents. The retriever component can be configured to supply the reader with the latest documents pertinent to the user's query, addressing the challenge posed by the LLMs' knowledge cut-off. Moreover, providing pertinent context helps reduce LLMs' tendency to hallucinate~\cite{shuster-etal-2021-retrieval-augmentation}.

Several enhancements can be considered to further increase the accuracy of  RAG. We explored two such enhancements: end-to-end training and query rewriting. Below, we outline these enhancements and explain our rationale for their inclusion or exclusion.

End-to-end training involves jointly training the retriever and the reader on domain-specific data~\cite{lewisrag, izacard2022atlas}. Previous attempts at end-to-end training have employed LLM readers such as BERT \cite{devlin2018bert} and BART \cite{lewis2019bart}. However, implementing this approach with readers like Llama 2~\cite{touvron2023llama} and GPT-3~\cite{brown2020language} remains prohibitively expensive. Since our chatbot is based on Llama 2, we do not pursue end-to-end training. Furthermore, we note that while end-to-end training has been shown to lead to improvements in studies with BERT and BART, these improvements are comparatively modest~\cite{lewisrag}. This suggests that end-to-end training would be worthwhile only if its cost is sufficiently low, which is currently not the case for the newer generation of LLMs.

A second possible enhancement to consider is query rewriting. In a RAG pipeline, the retriever step fetches documents similar to the user query. A well-written and self-contained query is thus critical for this step. Nonetheless, real-world user queries are not always optimal and may require adjustments to improve the retriever's accuracy.  Ma et al.~\cite{ma2023query} demonstrate the usefulness of adding a query rewriter to RAG. Motivated by their results, we adapt and extend their guidelines to integrate a query rewriter into our chatbot. Although query rewriting inevitably increases QA execution time, our overall pipeline's execution time remains acceptable, as we show in Section~\ref{sec:evaluation}.

\section{Related Work}\label{sec:related}
In software engineering, chatbots are increasingly used to assist with tasks such as code understanding, code generation, and quality assurance~\cite{Abdellatif2022}. 
This section reviews recent relevant strands of work on chatbot-enabled software engineering and contrasts them with our research.

A prominent example of a widely used conversational software development tool is GitHub Copilot~\cite{Copilot}. Copilot builds on top of the Codex model~\cite{Codex}, a fine-tuned version of GPT-3~\cite{brown2020language}, to provide support in various tasks such as code completion, code generation from natural-language input, code migration, and answering coding questions. Within Copilot, the task most similar to our chatbot function is answering coding questions. Nevertheless, Copilot’s question-answering capabilities are focused on code-related queries~\cite{CopilotChat}. In contrast, our chatbot does not have a code-centric focus and is complementary; its primary goal is to assist systems engineers with domain-specific technical questions concerning integration, testing, deployment, and troubleshooting.

QAssist~\cite{ezzini2023aibased} employs a RAG architecture similar to ours to help stakeholders analyze and improve the quality of software requirements specifications. Specifically, this tool uses requirements-relevant content alongside generic domain material sourced from Wikipedia to answer questions about natural-language requirements. While QAssist uses RAG and shares a broad-spectrum question-answering objective similar to ours, it implements extractive QA using BERT variants. As such, QAssist can only highlight passages containing the answer, without the ability to engage with users in a conversational and context-aware manner. In contrast, our chatbot uses Llama~2, a generative LLM, to produce coherent and context-aware answers. These answers are based on knowledge extracted from relevant passages and chat history. Furthermore, in terms of design, our chatbot has a more advanced technology chain than QAssist, aligning with the latest \hbox{advances in chatbot development.}

A recent study by Abedu et al.~\cite{abedu2024softwareRepositoryMining} employs a RAG-based chatbot to facilitate user access to information within software repositories. This approach allows users to input the URL of the target repository in the chatbot interface and then query the repository's content. While the ability to dynamically input the desired repository provides flexibility, it also complicates the implementation of tailored preprocessing to enhance retrieval performance. In our chatbot design, we separate the domain-corpus creation process from the chatbot pipeline. This separation enables us to preprocess documents based on their content type. Furthermore, while Abedu et al. only experiment with a fixed choice of retriever (embeddings-based), we conduct a comparative analysis of four different types of retrievers to identify the most suitable one for our problem context, as we discuss in Section~\ref{subsec:procedure}.

In summary, while chatbots have been deployed to support various software engineering tasks, none are specifically designed for the analytical goals that our chatbot addresses, nor do they have the exact same design considerations as ours.

%% file: chatbot.tex
\section{Chatbot Development}\label{sec:approach}
In this section, we describe the approach implemented by our chatbot. The chatbot takes as input a collection of documents -- in the context of our industry collaboration, a collection of Ericsson documents related to CI/CD -- alongside a natural-language user query. Subsequently, the chatbot uses an LLM to generate a natural-language response to the query. Section~\ref{subsec:indexing} describes the process of creating a domain-specific corpus from Ericsson's CI/CD documents. Section~\ref{subsec:chatbot architecture} provides an overview of our chatbot's architecture and discusses the steps involved in generating an answer based on the domain-specific corpus and a given query.

\begin{figure}
    \centering 
    \includegraphics[width=\linewidth]{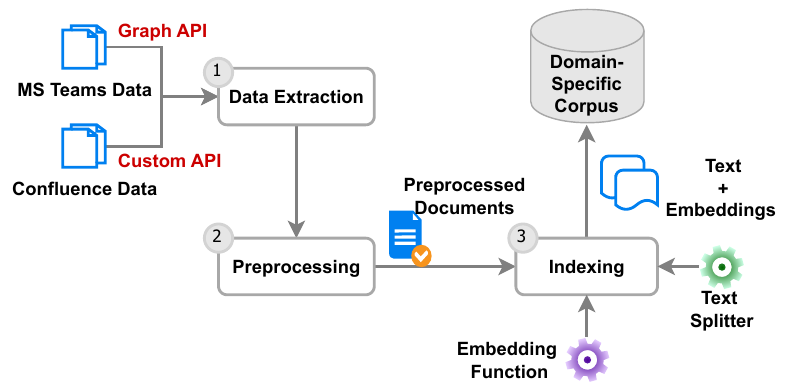}
    \caption{Steps for Creating a (Domain-specific) CI/CD Corpus}
    \label{fig:indexing}
\end{figure}

\subsection{Corpus Creation}\label{subsec:indexing}
Figure~\ref{fig:indexing} presents the steps we follow for transforming Ericsson's CI/CD documents into a (domain-specific) corpus.

\sectopic{Step 1: Data Extraction.} We gather CI/CD domain knowledge from two Ericsson-specific sources: (1) messages exchanged among software engineers in CI/CD channels on internal Microsoft Teams, and (2) Confluence~\cite{confluence} web pages that contain information about common troubleshooting procedures for CI/CD-related tasks within the organization. These data sources enable our chatbot to respond to Ericsson-specific technical questions. Noting that both data sources evolve frequently over time, we first extract the most recent data from these sources. This extraction step is decoupled from the chatbot itself (Section~\ref{subsec:chatbot architecture}), allowing updates to be performed offline and on a regular basis (e.g., overnight) without disrupting the chatbot's function.

The Teams data, comprised of messages and replies, is collected through the Microsoft Graph REST API~\cite{graphAPI}. Using this API, we generate separate tables (in CSV format) for storing messages and replies, with one table dedicated to messages and another to replies. The messages table has 35 parameters (fixed columns) such as message ID, creation time, content, sender's username, mentions, and reactions. The replies table consists of 43 parameters such as reply ID, reply's parent message ID, content, mentions, and reactions. \comm{A complete list of parameters for Microsoft Teams messages and replies is provided in Appendices \ref{app:messages} and \ref{app:replies}, respectively.}

The Confluence data consists of a set of HTML pages, with each page containing a troubleshooting topic followed by content outlining the troubleshooting procedure. We extract these pages using a custom REST API developed by Ericsson.

\sectopic{Step 2: Preprocessing.}
We process the data extracted in Step~1 into a suitable format for use by the LLM. 

For the Teams data, we strip the HTML formatting from the `content' column of the messages and replies tables discussed earlier and convert the content to plain text. Subsequently, each reply in the replies table is linked to its corresponding parent message in the messages table through the unique parent message ID. This process connects each message to its replies in the same order as they were originally posted, thus preserving the chronological order of messages. To protect privacy, as we reconcile the messages and the responses to them, we remove personal information such as employee names and email addresses from both the content and the associated metadata. Finally, we store every message along with all the responses to it in a plain-text document.

As for the extracted Confluence pages, the preprocessing is straightforward: we store the title and content of each page in a plain-text document.

To facilitate more accurate interpretation of the Teams and Confluence data by the LLM, we augment this data with prefixes. Specifically, each Teams message is prefixed with the phrase ``Message:'', while replies to the message are prefixed with ``This message has the following responses:''. Similarly, the title of each Confluence page is prefixed with ``Page Title:'', and the content following the title is prefixed with ``The content of this page is as follows:''.

\sectopic{Step 3: Indexing.}
In this step, we embed and store the preprocessed documents obtained from Step~2 (Figure~\ref{fig:indexing}) in a vector database. This vector database serves as the domain-specific corpus for the information retrieval step of our chatbot (Step~2 in Figure~\ref{fig:chatbot-design}, discussed later). Since the preprocessed documents are ultimately supplied to the LLM as relevant context, our objective is to maximize their length. However, two important considerations arise when determining the ideal length. On the one hand, we must ensure that the documents fit within the context length (token limit) accepted by the LLM, as exceeding this limit could result in context loss, runtime errors, or incoherent output. On the other hand, supplying long documents to the LLM can lead to a ``needle in the haystack'' problem~\cite{hassani2024rethinking}, where relevant information is lost amongst the noise. Therefore, determining the optimal length of embedded documents becomes an important factor in maximizing chatbot efficacy, as we aim to maximize the amount of relevant information while limiting the noise.

Our choice of how to split the data in a preprocessed document depends on the source from which the document originates. We have custom splitters for each Teams and Confluence. For Teams, based on actual data and the experience of collaborating engineers at Ericsson, the combination of a message and all its replies is anticipated to always be well below the contextual length of modern LLMs. Therefore, we have determined that each individual message, alongside all the replies to it, could be embedded directly as one unit in our vector database. This means that the units fetched by the retriever step of the chatbot will constitute one message and all the associated replies.

For the Confluence data, the length of a preprocessed document could exceed the context length of the LLM. Therefore, we need to apply document chunking before embedding large Confluence documents in the vector database. To determine the optimal chunk size for Confluence data, we conducted exploratory experimentation with values ranging from 200 to 1000 tokens (increasing the chunk size by 100 tokens in each iteration). Based on this experimentation, we observed that a chunk size of 800 tokens led to the best question-answering results. Furthermore, we followed the best practice of making adjacent chunks overlapping~\cite{ezzini2023aibased}, maintaining an overlap of 200 tokens (25\%) between adjacent chunks. To further preserve context, all chunks belonging to the same Confluence document were prefixed with the document title and the respective chunk number.

Following the splitting of the Teams and Confluence data, we send the resulting chunks to an embedding function to generate text embeddings. To ensure consistent terminology, we refer to these chunks as \emph{context items} rather than ``documents'', to avoid ambiguity between the chunks and the original (Confluence) documents. We store the embeddings for each context item in the vector database alongside the item's original text. The domain-specific corpus depicted in Figures~\ref{fig:indexing} and \ref{fig:approach} is realized by this vector database.

\subsection{Chatbot Design}\label{subsec:chatbot architecture}
\begin{figure}
    \centering
    \includegraphics[width=\linewidth]{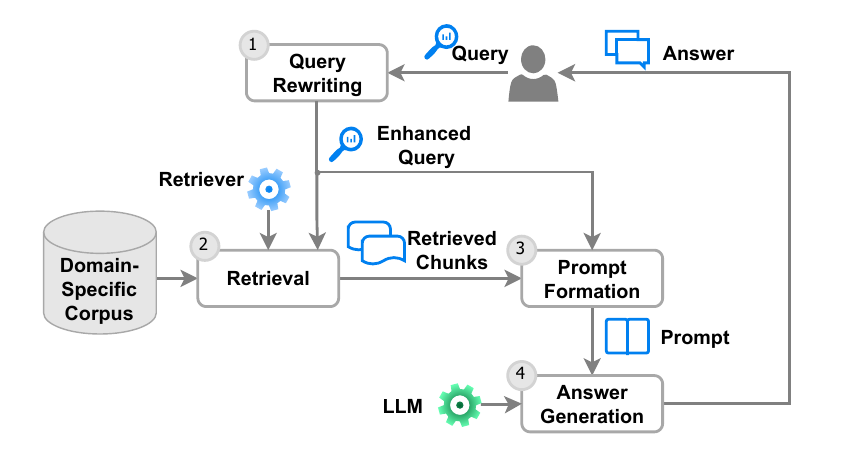}
    \vspace*{-1.5em}
    \caption{Overview of Our Chatbot Design}\label{fig:chatbot-design}
    \label{fig:approach}
    \vspace*{-.5em}
\end{figure}

Figure~\ref{fig:approach} shows the steps implemented in our chatbot. We will discuss each of these steps below.

\sectopic{Step 1: Query Rewriting.}
Given that user queries may not always be clear or well-structured~\cite{ma2023query}, our query rewriting module prompts an LLM\,\footnote{Note that the LLM used for query rewriting does not necessarily have to be the same one used for answer generation in Step~4 of Figure \ref{fig:approach}, which will be discussed later. In our current approach, nevertheless, we use Llama~2 as the LLM of choice to implement both Steps 1 and 4 of our chatbot design.} to enhance the query into a more effective search query. To achieve this, we use the query enhancement guidelines of Ma et al.'s~\cite{ma2023query}. Specifically, given a (frozen) LLM and a user query, we prompt the LLM to restate the query in more precise terms before the query is used for question answering. We augment Ma et al.'s prompt template for query rewriting by instructing the model to also analyze the user query in the context of the conversation history and determine if the current query is a follow-up. 
In response, the LLM either forms an improved question or returns the user query \hbox{verbatim as the question to pose to the LLM.}

\begin{figure}[!t]
    \centering
    \includegraphics[width=\linewidth]{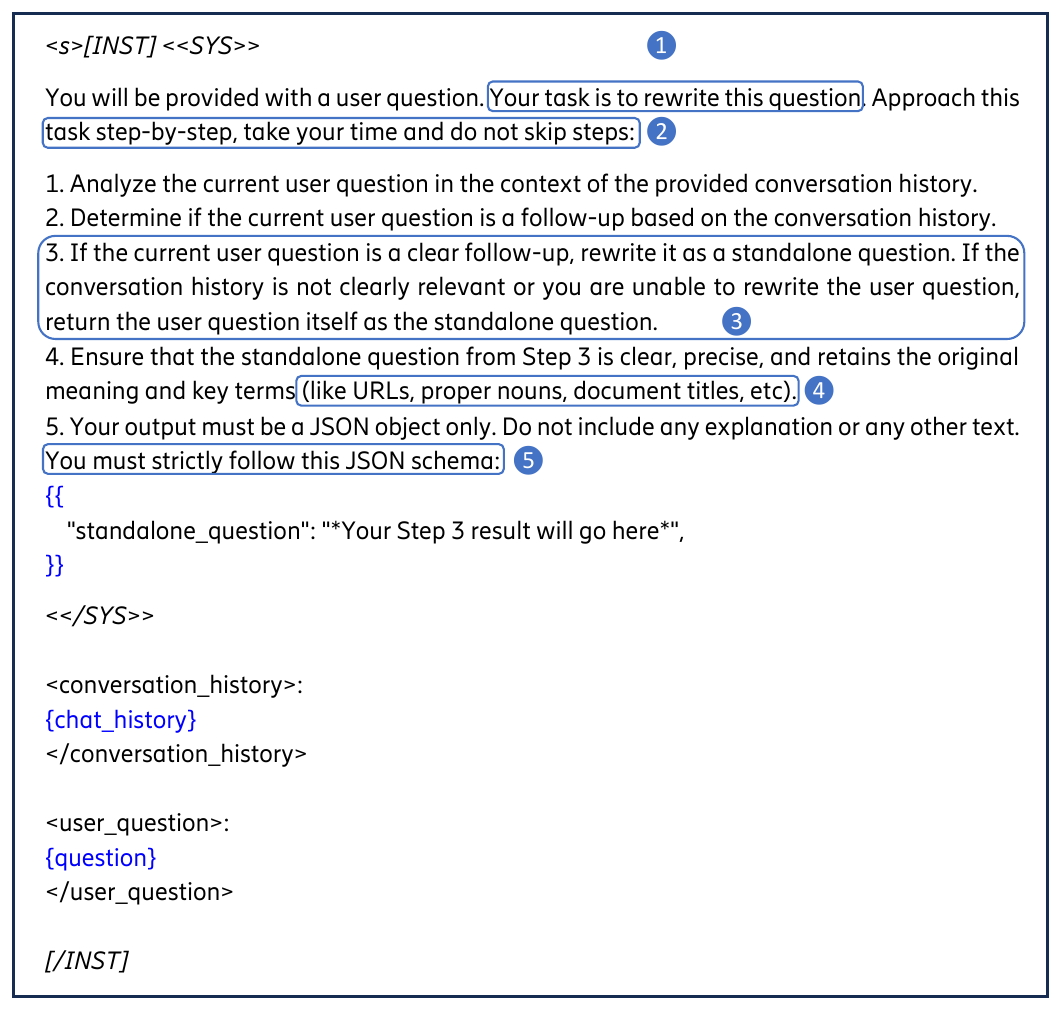}
    \vspace*{-1.5em}
    \caption{Prompt Template for Query Rewriting}
    \label{fig:query rewriting prompt}
    \vspace*{-.5em}
\end{figure}

In addition to attempting to construct a better query based on Ma et el.'s prompt template, we incorporate various prompt engineering techniques to improve the quality of the LLM's output. Figure~\ref{fig:query rewriting prompt} presents our query-rewriting prompt  and highlights the various considerations involved. 

We employ the same prompt template that Meta utilized for pre-training Llama~2~\cite{touvron2023llama}. This helps ensure that the model is able to understand the prompt structure and instructions clearly. In the prompt template, the instructions provided between the \verb|<<SYS>>| tokens convey a system message to the model, instructing it on its task and intended behaviour. To improve results, we explicitly  define the task of the LLM (Figure~\ref{fig:query rewriting prompt}, \StepOne). We then employ zero-shot chain-of-thought prompting~\cite{wei2022chainofthought} (\StepTwo) to decompose the task into intermediate steps. We mitigate ambiguity in the model's understanding of the task by explicitly handling possible scenarios  (\StepThree). Further, we instruct the model to retain important key terms from the query (\StepFour) and to strictly follow the desired output format (\StepFive).

\begin{figure}[!t]
    \centering 
    \includegraphics[width=\linewidth]{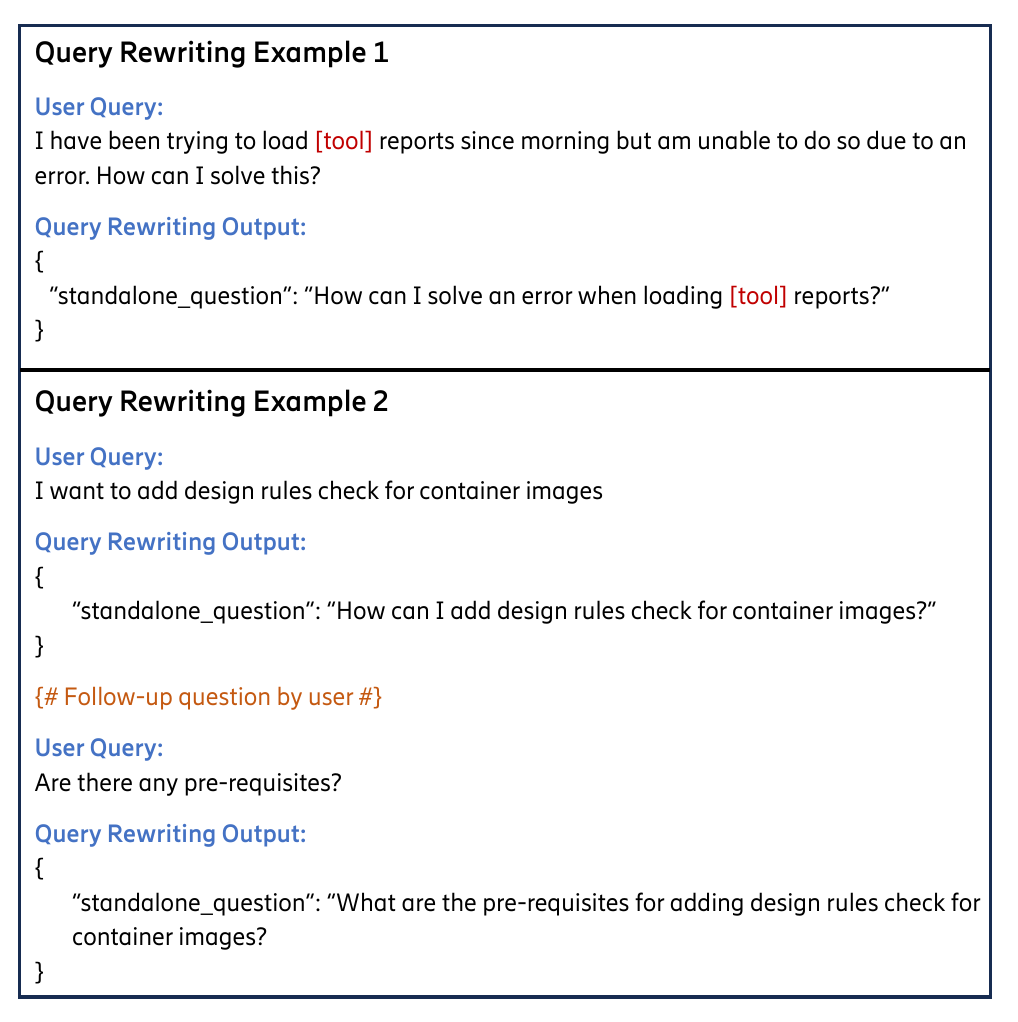}
    \vspace*{-2em}
    \caption{Examples of Query Rewriting}
    \label{fig:Query Rewriting Examples}
    \vspace*{-1em}
\end{figure}

We illustrate Step~1 using the two examples, Example~1 and Example~2, in Figure~\ref{fig:Query Rewriting Examples}. Example~1 demonstrates the ability of query rewriting to filter irrelevant information from the user query and generate a better query that focuses on the important aspects. Example~2 highlights how query rewriting can identify the question implied by the user statement. Moreover, in the case of the follow-up question in Example~2, query rewriting uses the conversation history to derive a complete query.

\sectopic{Step 2: Retrieval.}
Following the retrieval-augmented generation approach~\cite{lewisrag}, we employ a retriever component to identify domain-specific knowledge that should be imparted to the LLM for answer generation. In our design, the retriever uses the enhanced query obtained from query rewriting (Step~1) to retrieve relevant context items from the domain-specific corpus, constructed using the process \hbox{described in Section \ref{subsec:indexing}.}

Recognizing that different data sources may necessitate different retrievers for better question-answering accuracy, our approach offers the flexibility to specify the retriever to be used. The selected retriever is responsible for picking the top-$k$ context items to feed to the LLM as relevant information for answer generation. Determining the optimal value of $k$ is crucial and should consider the nature and size of the context items. Our exploratory experimentation revealed that setting $k=3$ yields the best results in our application context. This choice also happens to be consistent with the $k$ value recommended by Ezzini et al.~\cite{ezzini2023aibased} based on systematic experimentation. However, we observe that forcing $k$ context items to be considered at all times has the potential to introduce noise in cases where the number of relevant items in the corpus is less than $k$. To filter irrelevant information in the top-$k$ context items, we consider a context item only if it has a higher cosine similarity value to the query than a configurable threshold. This threshold helps prevent the selection of irrelevant context items merely to meet the specified quota of $k$ items. We set this threshold to 0.7, based on exploratory experimentation and informed by our experience in setting similar thresholds in our previous work~\cite{luitel2024improving}.

\begin{figure}
    \centering
    \includegraphics[width=\linewidth]{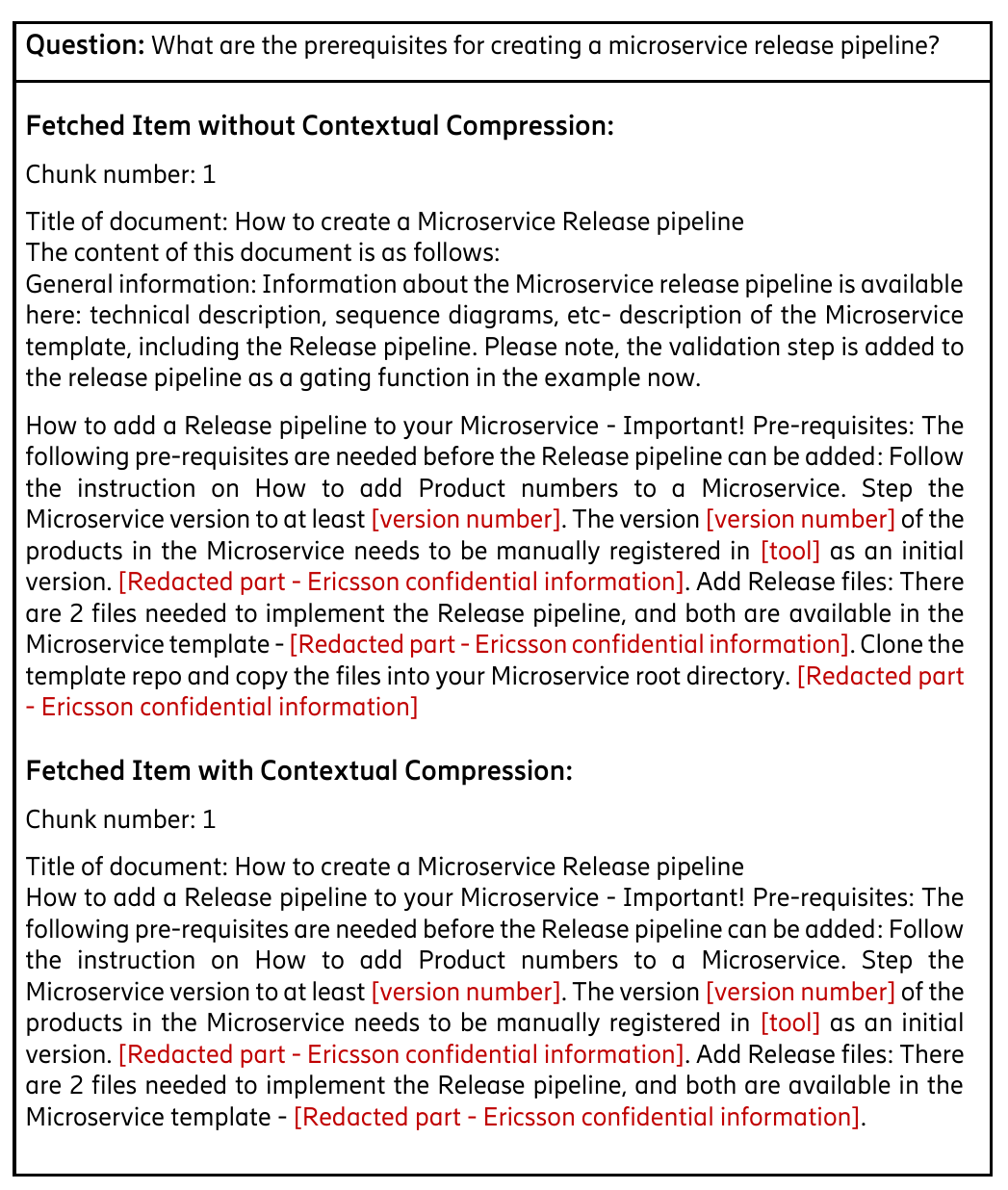}
    \vspace*{-.5em}
    \caption{Example of Contextual Compression}
    \label{fig:Contextual Compression Example}
    \vspace*{-.5em}
\end{figure}

Finally, we employ reordering and contextual compression techniques to further enhance the quality of the retrieved context items. The reordering technique involves placing the most relevant context items at either the beginning or the end when feeding the material to the LLM. This approach was inspired by recent research, which indicates that LLMs utilize context most effectively when it is located at the beginning or end, with a decline in performance when the relevant context is situated in the middle of long contexts \cite{liu2023lost}. Moreover, since the information relevant to the user query might be buried within the fetched items, we attempt to compress the items using the query before feeding the items to the LLM. This process, known as contextual compression~\cite{contextualcompression}, helps reduce the amount of irrelevant information. Similar to the threshold discussed above, contextual compression is a noise-reduction measure; however, whereas the threshold filters out irrelevant items, contextual compression mitigates the noise present within the selected items. Figure~\ref{fig:Contextual Compression Example} presents an example of contextual compression, illustrating that even though the retrieved item remains the same in both cases, contextual compression significantly reduces the noise present within the item with respect to the question.

\begin{figure}
    \centering
    \includegraphics[width=\linewidth]{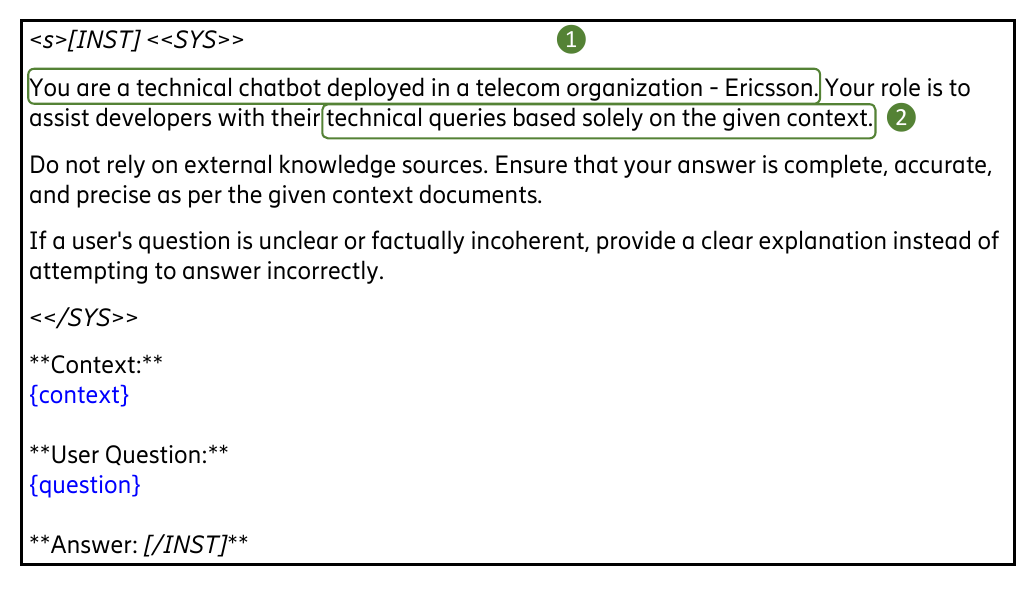}
        \vspace*{-2em}
    \caption{Prompt Template for Answer Generation}
    \label{fig:answer generation prompt template}
        \vspace*{-.5em}
\end{figure}

\sectopic{Step 3: Prompt Formation.} 
This step takes as input the enhanced query generated from Step~1 and the relevant items retrieved in Step~2, and formulates a question prompt for the LLM. Figure~\ref{fig:answer generation prompt template} presents our prompt template. In this template, we clearly define the environment in which the model is being deployed, as well as its responsibilities (Figure~\ref{fig:answer generation prompt template}, \StepOneAnswerGeneration). This guides the model to refrain from answering unethical questions. Furthermore, the prompt includes a safeguard to mitigate model hallucinations. Specifically, we instruct the model to answer the query solely based on the retrieved documents (\StepTwoAnswerGeneration). In cases where the model is unable to provide an accurate answer, it is instructed to inform the user and either ask for more clarity (if the query is still unclear) or request more information (if sufficient relevant context has not been retrieved). The subsequent portion of the template contains placeholders for the context items and the user question, which are populated at runtime. After these placeholders are populated, the final prompt is passed to the LLM.

\sectopic{Step 4: Answer Generation.} 
\begin{figure}
\centering
\includegraphics[width=\linewidth]{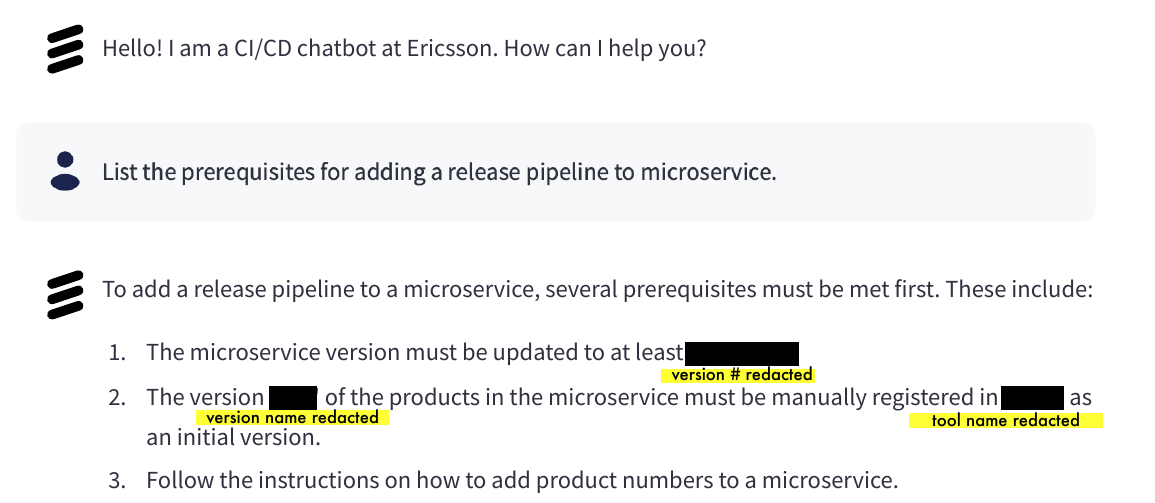}
    \caption{Example of a User Query and Response Interaction}
    \label{fig:web application}
        \vspace*{-.5em}
\end{figure}
In this step, we use the LLM, which, in our current implementation, is Llama~2, to generate a response to the prompt obtained in Step 3. Figure~\ref{fig:web application} illustrates an example response by the chatbot to the following query: ``List the prerequisites for adding a release pipeline to microservice'', which is a variant of the query shown earlier in Figure \ref{fig:Contextual Compression Example}. The chatbot takes this query as input, retrieves relevant context items, and generates \hbox{an answer based on them.}

%% file: exp.tex
\section{Empirical Evaluation}\label{sec:evaluation}
\subsection{Research Questions}
Our evaluation aims to answer the following two research questions (RQs):
\par\textbf{RQ1.} \emph{How accurate is our chatbot?} RQ1 assesses the accuracy of our chatbot using a combination of automated metrics and a manual analysis of correctness. This manual analysis is followed by a root-cause analysis, which identifies the underlying reasons for the inaccuracies in the chatbot's responses as observed in our case study.

\par\textbf{RQ2.} \comm{\emph{What is our chatbot's response time?}} RQ2 measures the execution time of different chatbot-pipeline instantiations.

\subsection{Implementation}
We implement our chatbot using Python (version 3.10) along with supporting libraries. Specifically, we utilize the Transformers library (version 4.31.0)~\cite{transformers} for loading the model and the language tokenizer. To reduce computational requirements, we load the model in a 4-bit quantized format using the BitsAndBytes library (version 0.41.0)~\cite{bitsandbytes}. HuggingFace serves as the model repository and provides a wrapper for the text-generation pipeline over the model.

For our experiments, we use the Llama2-chat 7B parameter model~\cite{touvron2023llama} released by Meta. Llama2-chat is an open-source model, and by hosting it locally, we ensure that Ericsson's confidential data remains secure. Furthermore, this model has been optimized for and has demonstrated strong performance in conversational applications~\cite{touvron2023llama}. Finally, the model is compliant with Ericsson's internal LLM usage policies.

We employ the BAAI/bge-base-en embedding model~\cite{BAAIembeddingmodel} for indexing documents as it has shown good performance for the retrieval task and is computationally inexpensive~\cite{MTEBleaderboard}. LangChain (0.0.240)~\cite{Langchain} acts as the primary library, providing support and wrappers for the RAG functions. These include: (a) a wrapper over the Chroma embedding vectorstore~\cite{Chroma}, (b)~support for implementing the chat history window, (c)~a wrapper over the retriever, (d)~support for various optimizations of the retrieval process, and (e) a wrapper for the question-answer pipeline. We employ the Ragas library (version 0.0.21) \cite{RagasLibrary} for our evaluation process. 

\subsection{Experimental Setup}\label{subsec:experimental setup}
We deployed our chatbot and performed our experiments on a Kubernetes pod containing an Intel Xeon Gold 6230N CPU with 40 GB of RAM and an Nvidia Tesla T4 GPU with 16 GB of GDDR6 memory. The CUDA version was 12.4, and the OS used was Ubuntu 20.04. Model serving was performed using the RayServe software library~\cite{RayServe}.

\subsection{Ground Truth}
The ground truth for our evaluation consists of 72 representative question-answer pairs gleaned over a span of nearly a year from resources and internal communications among members of the CloudRAN team at Ericsson. For each question-answer pair, we recorded the basis (documents and/or messages) upon which the correct answer was articulated.

\comm{\subsection{Domain Corpus}
To build the domain corpus used as input for the retrieval step of our chatbot (Step 2 in Figure~\ref{fig:chatbot-design}), we followed the corpus creation process described in Section~\ref{subsec:indexing}. The corpus, which forms the basis for our evaluation in this section, was generated in May 2024. It includes 4,169 Microsoft Teams messages, 18,389 responses to these messages, and 240 Confluence web pages. These CI/CD resources collectively resulted in a total of 4985 context items, which were then stored in our vector database along with their embeddings.}

\subsection{Metrics} \label{subsec:metrics}
\subsubsection{Metrics for RQ1} To address RQ1, we evaluate the performance of the chatbot's retrieval component through context recall and assess the overall chatbot pipeline using answer similarity, as explained below.

\textbf{Context Recall@k}  measures whether the correct answer to a user question is present within the top $k$ context items fetched by a retriever. In other words, given a question and $k$ retrieved context items, this metric checks if one of these items contains all or part of the answer to the question.
    
\textbf{Answer Similarity} evaluates the semantic resemblance between the generated answer and the ground-truth answer. The generated and ground-truth answers are first embedded using an embedding function to create vectors. We employ the same embedding function that was used in Section~\ref{subsec:indexing} (Step~3) to create the document corpus. The semantic association between these vectors is then determined using cosine similarity. The metric value ranges between 0 and 1, with a higher value indicating greater resemblance and, thus, better accuracy.

In addition to the above automatically computed metrics, we manually evaluate a single iteration of all ground-truth questions answered by our best-performing pipeline (as per Recall@k and answer similarity), classifying the chatbot answers as correct, partially correct, or incorrect:

\textbf{Correct.} A (generated) answer is correct if it is semantically equivalent to the ground truth. An answer is deemed equivalent to the ground truth if (a) it does not omit any information present in the ground truth, \emph{and} (b) it does not include any information absent from the ground truth.

\textbf{Partially Correct.}  An answer is classified as partially correct if it (a) includes extraneous information not present in the ground truth, (b) is incomplete, i.e. missing information that is in the ground truth, \emph{or} (c) is both incomplete and contains extraneous information.  

\textbf{Incorrect.} An answer is incorrect if it has no content intersection with the ground truth.

\vspace*{.5em}

\subsubsection{Metric for RQ2} To address RQ2, we measure the chatbot's response time in seconds, defined as the duration from when the user submits a question to when the chatbot provides the full answer. The response time was computed over the experimental setup described in Section~\ref{subsec:experimental setup}. A basic measure of the usefulness of the chatbot would be for its response time to be less than that of a human expert answering the same question through messaging channels, e.g., on Teams.

\subsection{Evaluation Procedure}\label{subsec:procedure}
Since our choice of LLM for answer generation is restricted to Llama~2 following Ericsson's security guidelines, our evaluation procedure is focused on assessing the performance of this particular LLM when combined with various retrievers. We examine the following four retrievers in our study:

\textit{(a) TF-IDF-based retriever.} TF-IDF evaluates the importance of terms within a document relative to a corpus -- in our context, the corpus created as described in Section~\ref{subsec:indexing} -- by considering both term frequency (TF) and inverse document frequency (IDF). Specifically, this retriever ranks context items against the user query based on the combined weight of term frequency in the item and rarity across the entire corpus, effectively identifying context items that contain frequently occurring important terms \cite{TF-IDF}.

\textit{(b) BM25-based retriever.} BM25 is a probabilistic information retrieval method that attempts to overcome the drawbacks of TF-IDF by document length normalization~\cite{BM25}. This normalization allows BM25 to account for varying document lengths and to prevent longer documents from having an unfair advantage in the retrieval process. This retriever, like the TF-IDF retriever, employs a domain corpus to identify the top-$k$ relevant items to the user query.

\textit{(c) Embeddings-based retriever.} This retriever uses the embedding of the user query and the embeddings stored in the domain corpus (Section~\ref{subsec:indexing}) to retrieve the top-$k$ relevant items. The retrieval process has three main steps. First, the retriever embeds the query using the same embedding function as that used for embedding the domain-specific corpus during the indexing process (Step~3 of Figure~\ref{fig:indexing}). Second, the retriever computes cosine similarity scores between the embedded query and the items in the vector database. Finally, the retriever selects and returns the top-$k$ most semantically similar items.

\textit{(d) Ensemble retriever.} This retriever calculates the simple average of the scores from the BM25- and embeddings-based retrievers and returns the top-$k$\,items according to the averages.

To prepare the evaluation dataset for each chatbot pipeline instantiated with a different retriever, we iterate over the ground-truth questions and store the chatbot's answer along with the retrieved context items. The final evaluation dataset includes ground-truth questions and answers, the context item(s) containing the correct answer, the chatbot's generated answer, and the retrieved context item(s). 

We compute the Recall@k and answer similarity metrics for each retriever using its respective dataset. To account for variations in model output, we conduct three evaluation iterations for each retriever and report averages for answer similarity, Recall@k, and response time.
  
Using the Recall@k and answer similarity results, we determine the chatbot pipeline that has the best accuracy. After identifying the most accurate pipeline, we manually analyze the answers generated by one run of this pipeline over all the questions in the ground truth. 
Through this analysis, we categorize the generated answers as correct, partially correct, or incorrect, as defined in Section~\ref{subsec:metrics}. We then conduct a qualitative error analysis to better understand the inaccuracies in the generated answers.

\subsection{Answers to RQs}

\begin{table}[!t]
    \caption{Automatically Computed Accuracy Results}
    \label{tab:performance metrics}
    \centering
    \begin{tabular}{|l|c|c|c|c|c|}
        \hline
        \textbf{Metric} & \textbf{TF-IDF} & \textbf{BM25} & \textbf{Embedding} & \textbf{Ensemble} \\
        \hline
        Recall@3 & 91.60\% & 91.60\% & 92.75\% & 95.10\% \\
        \hline
        Answer Similarity & 93.50\% & 94.40\% & 94.30\% & 95.40\% \\
        \hline
    \end{tabular}
    \vspace*{-.5em}
\end{table}

\textbf{RQ1. How accurate is our chatbot?}
\comm{We answer this question using the metrics defined in Section~\ref{subsec:metrics}.}
Table~\ref{tab:performance metrics} presents the Recall@k and answer similarity scores with  k~$=3$. \comm{The rationale for selecting this specific value of k was discussed in Section~\ref{subsec:chatbot architecture}.} As seen from the table, our pipeline performs well when instantiated with any one of the four retrievers. All pipelines fetch the correct context in more than 90\% of the cases. The term-based TF-IDF and BM25 retrievers achieve virtually the same Recall@3 results, with an average of 91.60\%. The embeddings-based retriever performs slightly better, with an average Recall@3 of 92.75\%. \emph{\bfseries The ensemble retriever yields the best overall results, with Recall@3 averaging at 95.10\%, slightly outperforming the TF-IDF, BM25 and embeddings-based retrievers by \comm{margins} of 3.5\%, 3.5\%, and 2.35\%, respectively.}

As for answer similarity, we get comparable results for all retrievers with a difference of $<$2\% between the different pipeline instances. 
Similar to Recall@3, the ensemble retriever-based pipeline yields the best performance for answer similarity, achieving an average score of 95.10\%.    

Since the ensemble-retriever-based pipeline had the best performance across both Recall@3 and answer similarity metrics, we select it for the subsequent manual error analysis. 

\vspace*{.2em}

 Table~\ref{tab:manual analysis} summarizes our error analysis results. \emph{\bfseries Out of the 72 answers generated by the chatbot in response to the ground-truth questions, 44 (61.11\%) were correct, meaning that they were complete and did not contain any extraneous information. \comm{In all cases where the answer was correct, the retriever always retrieved the correct context item(s).} A total of 19 (26.39\%) answers were partially correct: 11 (15.28\%) were missing vital information, 3 (4.17\%) contained additional information that could be misleading, and 5 (6.94\%) were both incomplete and further contained orthogonal information. 
Finally, 9 (12.50\%) answers were deemed incorrect as they had no overlap with the \hbox{ground-truth answers.}}

\begin{table}
    \caption{Results of Manual Analysis}
    \label{tab:manual analysis}
    \centering
    \begin{tabular}{|c|c|c|c|c|}
        \hline
        \textbf{Correct} & \multicolumn{3}{c|}{\textbf{Partially Correct}} & \textbf{Incorrect} \\
        \cline{2-4}
        & \textbf{(A)} & \textbf{(B)} & \textbf{(C)} & \\
        \hline
        44 & 11 & 3 & 5 & 9 \\
        \hline
    \end{tabular}

    \vspace*{.5em}
    
    \parbox{.6\linewidth}{\textit{(A)~Only Incomplete, (B)~Only Extraneous, (C)~Both Incomplete and Extraneous}}
    \vspace*{-1em}
\end{table}

\begin{figure}
    \centering
    \includegraphics[width=\linewidth]{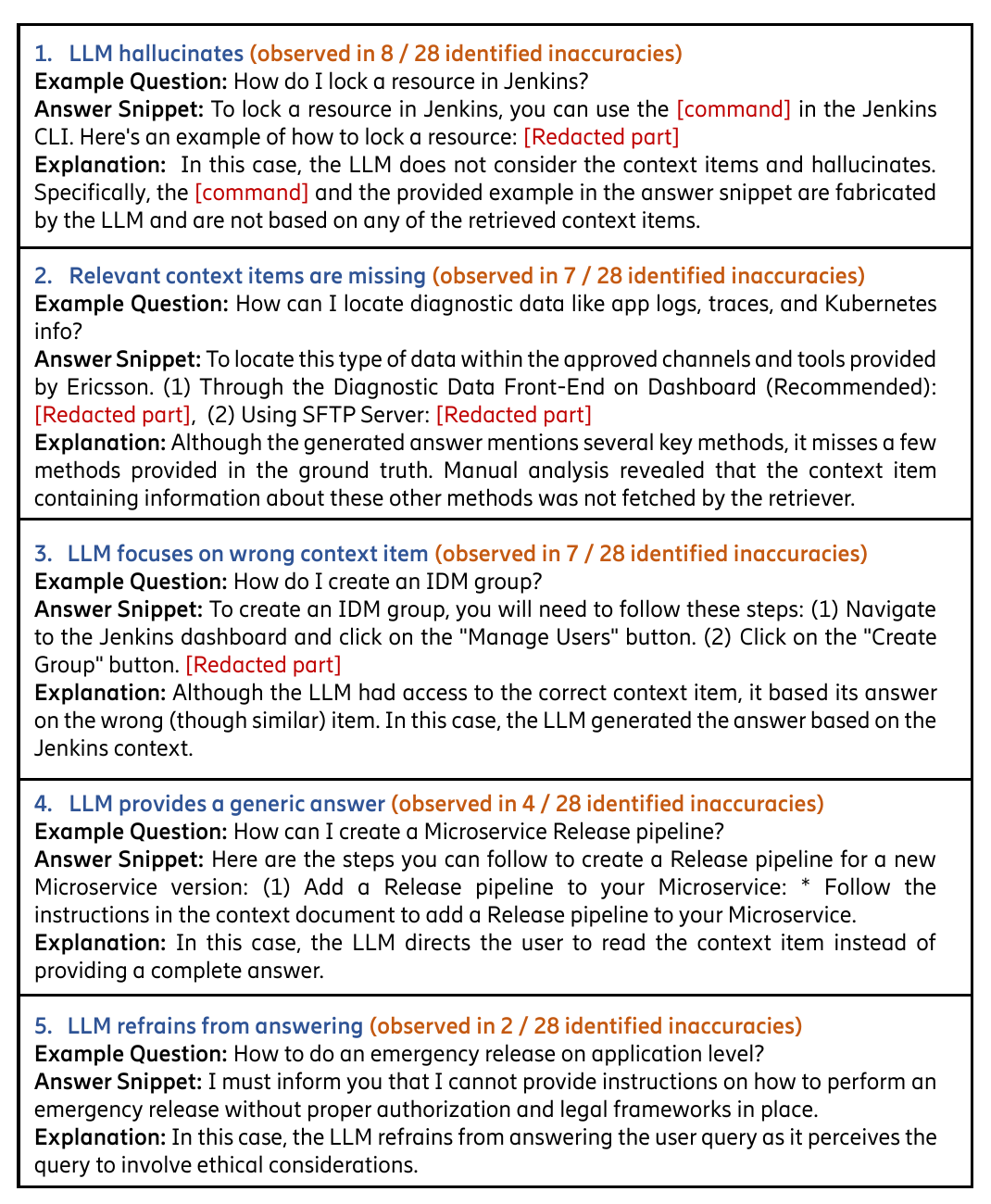}
    \vspace*{-1.5em}
    \caption{Results of Error Analysis, including Root Causes of Inaccuracies and Their Prevalence in Our Case Study}
    \label{fig:error causes}
    \vspace*{-1em}
\end{figure}

\emph{\bfseries Root causes of errors.} We analyzed the incorrect and partially correct answers to identify the root causes of errors. Figure~\ref{fig:error causes} presents the identified causes with examples and explanations.
Hallucinations were the most prevalent error, affecting 8 answers. The observed hallucinations  occurred despite the relevant context items being retrieved. These items were nonetheless ignored by the LLM, which then proceeded to generate its own answer. This indicates that the LLM does not always follow the answer-generation prompt (Figure~\ref{fig:answer generation prompt template}).

The second and third most major causes (tied in terms of the number of observations) were either that the retriever did not fetch the correct context items or that the LLM focused on the wrong context items. Each of these affected 7 answers. The absence of relevant context items was the most common reason for incomplete answers and occurred particularly for responses that spanned  multiple context items. The second category of errors occurred when, despite the retriever having retrieved the correct context item(s), the LLM derived an answer from the incorrect context item(s) that were retrieved alongside the correct one(s) by the retriever.

The LLM providing a generic answer or refraining from giving an answer were the other root causes identified, respectively affecting 4 and 2 answers. In the former case, the LLM's tendency to reply with generic answers~\cite{luitel2024improving} prevents it from utilizing the context to provide tailored responses. In the latter case, the LLM refrains from answering the question due to one of the following reasons: an incorrect perception that there are ethical considerations, an inability to understand the question, or insufficient context.

\begin{figure}
    \centering \mbox{\hspace*{-2em}}
    \includegraphics[width=1.0\linewidth]{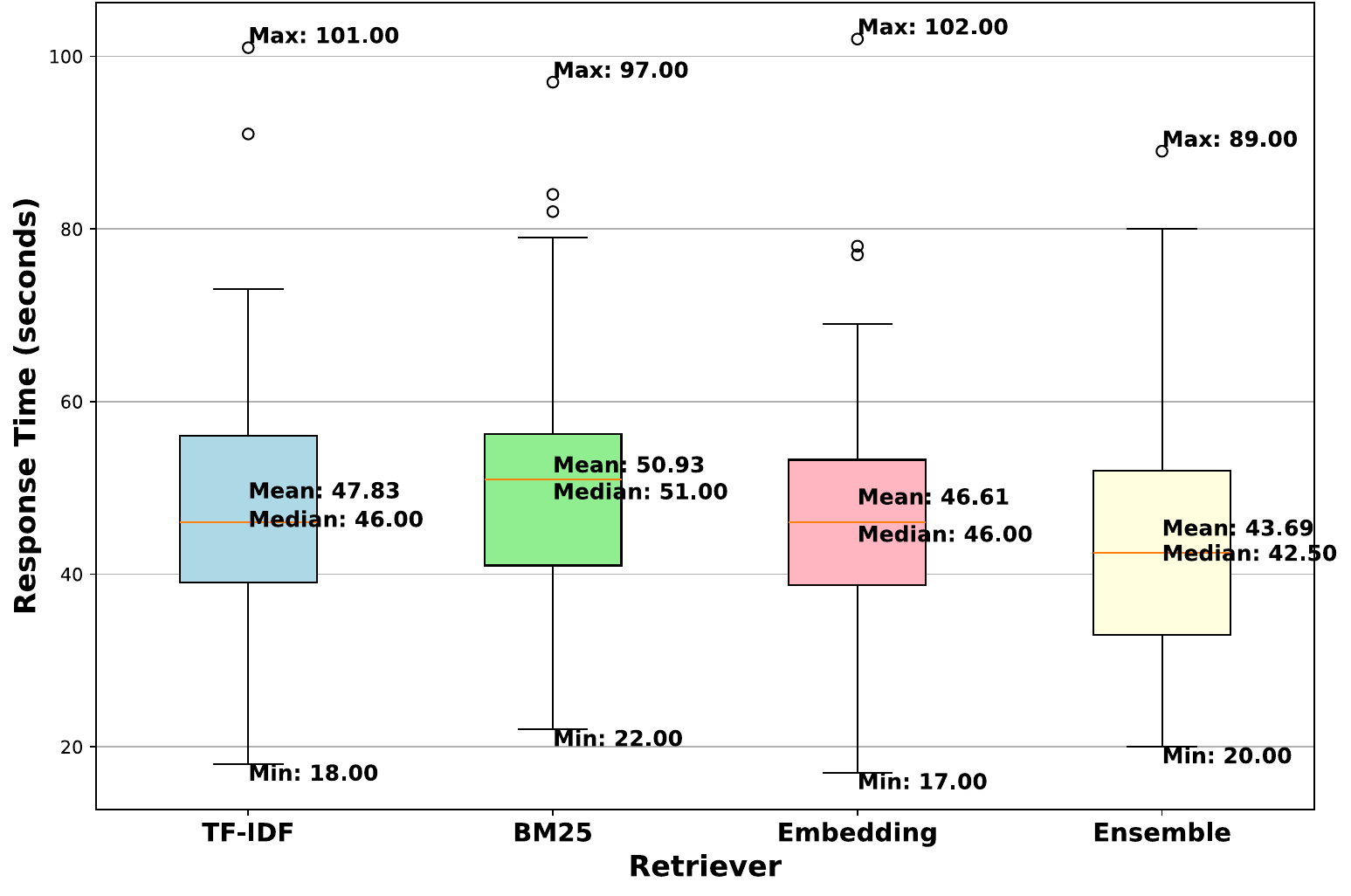}
    \vspace*{-.75em}
    \caption{Response Times for Chatbot Instances Using Different Retrievers}
    \label{fig:comparison of response times}
    \vspace*{-.5em}
\end{figure}

\comm{\textbf{RQ2. What is our chatbot's response time?}}
We show  in  Figure~\ref{fig:comparison of response times} the response times (in seconds) for the four chatbot pipelines induced by the four choices of the retriever component as discussed in Section~\ref{subsec:procedure}. Each boxplot in the figure represents the response times for one specific pipeline, with each data point being the response time for an individual question in the ground truth.
The TF-IDF- and BM25-based pipelines have average response times of 47.83 and 50.93 seconds respectively, with corresponding median values of 46 and 51 seconds. The embedding and ensemble pipelines had average response times of 46.61 and 43.69 seconds, with median values of 46 and 42.50 seconds respectively. Considering the modest hardware resources in our experimental setup (see Section~\ref{subsec:experimental setup}), these response times seem reasonable. Reductions in execution time should be relatively easy to achieve with improved hardware, such as multiple GPUs.

\comm{In our experiments, we found that the execution time is overwhelmingly dominated by the LLM during the query rewriting and answer generation phases (Steps 1 and 4 in Figure~\ref{fig:chatbot-design}). On average, these steps account for 99.91\% of the total execution time. In contrast, the retrieval and prompt formation steps (Steps 2 and 3) take negligible time, contributing less than one-tenth of a percent to the overall execution.  Based on the findings from RQ1, the ensemble-retriever-based pipeline has the highest accuracy. Given that the retrieval step has virtually no impact on execution time, and considering that the overall execution time of the ensemble-retriever-based pipeline is comparable to (or slightly better than) alternatives, as shown in Figure~\ref{fig:comparison of response times}, we conclude that the ensemble-retriever-based pipeline is the optimal choice for our chatbot.}

To assess our chatbot's usefulness, we analyzed the response times of experts to user queries in the Teams CI/CD channels at Ericsson, based on message timestamps. Our analysis revealed that the quickest response time from an expert in Teams was approximately 5 minutes.
Thus, while our chatbot could be faster, even on our modest hardware setup, it is sufficiently quick to be valuable to the querying party. \comm{That said, to conclusively determine if the response time is practical, a user study is needed, as a human interacting with a chatbot might expect a timely, synchronous conversation, whereas someone asking a question in a Teams channel with colleagues might find loosely asynchronous communication acceptable.}
Regardless of the speed of the chatbot's responses, there are  inherent time-saving benefits for whoever has to answer.

\subsection{Limitations and Validity Considerations}
\textit{\bfseries Limitations.} Our experiments focused exclusively on Llama~2 as the LLM of choice. This decision was driven by security protocols set by our industry partner, which restrict the use of alternative models on their proprietary data at this time. While further benchmarking with different LLMs remains important, our exclusive use of Llama~2 is unlikely to be a significant limitation, as Llama~2 is a state-of-the-art model that reflects current open-source LLM capabilities well.

\comm{We note that, without a user study, our current empirical results do not provide definitive evidence that our chatbot is ready for wide use at the host company, considering the inaccuracies observed and reported in RQ1. Nevertheless, we can make the following remarks which are likely to increase the likelihood of industrial adoption: First, the state-of-the-art in LLM technologies is evolving rapidly. We present a mature chatbot design that can be instantiated with newer LLMs. We anticipate that error rates will decrease further as more advanced LLMs become available, without our chatbot design being affected. Second, recognizing that chatbots are not infallible, engineers do not rely solely on chatbot responses for decision-making; they further consult with subject matter experts and use a range of other tools to guide their final decisions. These additional steps provide a safeguard against incorrect decisions stemming from inaccurate chatbot answers, thereby mitigating risks associated with using a chatbot that does not have perfect accuracy.}

\textit{\bfseries Threats to Validity.} The validity aspects most pertinent to our evaluation are internal, construct and external validity.
Regarding \emph{internal validity}, we note that data leakage poses a validity threat for LLM-based solutions if the model is exposed to test data during training. In our evaluation, the dataset is proprietary, and we can assert with reasonable confidence that Llama~2 was not exposed to this data during its pre-training.  Regarding \emph{construct validity}, we note that our accuracy metric for the retrieval step, Recall@k, aligns with empirical practices in the software engineering community~\cite{ezzini2023aibased}. To assess our chatbot's answer accuracy, we combine semantic similarity with manual human judgment. We opted for semantic similarity because it is more meaning-based than metrics like BLEU and ROUGE, which have been shown to have low correlation with human judgment~\cite{liu2016howNotToEval}. Developing suitable chatbot-evaluation metrics is an ongoing research topic~\cite{gao2024llmbased}. Given our detailed manual analysis and the close semantic similarity scores for the different chatbot pipelines evaluated, we do not anticipate major construct-validity threats due to our choices about metrics. Regarding \emph{external validity}, we acknowledge that our results are limited to a single case study. Although the industrial context of our work provides valuable insights, we recognize that generalizable conclusions cannot be drawn from a single case. Further case studies are necessary to explore broader applicability.

%% file: lesson.tex
\section{Lessons Learned}\label{sec:lessons}
Below, we discuss the lessons learned from developing our chatbot. We believe these lessons will be most useful for  researchers and practitioners interested in understanding the challenges and limitations of current chatbot technologies.

\textbf{Beware of hallucinations; balance context carefully.} Based on our error analysis summarized in Table~\ref{fig:error causes}, the top three issues accounting for nearly 80\% (22/28) of observed inaccuracies are hallucinations, missed context by the retriever, and discarded context by the LLM. Mitigating these issues requires steps to reduce hallucinations, improve retriever results, and ensure the LLM focuses on the correct retrieved items. However, it is important to note that inherent trade-offs exist here: simply increasing the amount of context to avoid missing information can exacerbate hallucinations or worsen the problem of the LLM focusing on incorrect context.

\textbf{Switching LLMs may require major adaptation effort.} While, theoretically, one should be able to switch LLM models to newer versions or alternative LLM technologies, this was not our experience. Prompting styles and guidelines vary between different models, and even between different versions of the same model family, leading to various issues when these models are updated. For instance, as we explored the possibility of using a different LLM, it became apparent that our query rewriting component might require a major reevaluation. An important takeaway for us was that until further harmonization efforts occur across LLMs for interoperability, significant effort may be necessary to change the underlying LLM or to upgrade the models.

\textbf{Preprocessing is key for increasing accuracy.} Preprocessing of documents requires careful consideration, as it can significantly impact chatbot accuracy. In our case study, initial testing showed suboptimal performance. Root-cause analysis revealed that despite maintaining an overlap to preserve contextual relationships, as suggested in the literature~\cite{ezzini2023aibased}, the retriever failed to fetch all subsequent items necessary to answer the questions. This insight led us to include the document title and chunk number in each context item, which resulted in major accuracy improvements. The lesson learned here is that such examinations and improvements in preprocessing, although often time-consuming, should be prioritized as they can have a drastic impact on the success of a chatbot. 

\textbf{Handling both domain-specific and general queries presents a challenge.} An important tradeoff exists between supporting general and specific queries in a RAG-based chatbot architecture like ours, where instructing the chatbot to respond exclusively based on information retrieved by the retriever component enhances accuracy by reducing hallucinations but, at the same time, limits the chatbot's ability to use its pre-trained knowledge. One could consider using a classifier to differentiate between domain-specific and general queries and direct different types of queries to different chatbots. However, we opted against this approach due to its potential complexity and error-proneness, as well as the undesirable consequences of misclassifying queries. In particular, we observed that distinguishing between a general query like ``What is the process to migrate from one Kubernetes cluster to another?'' and a specific one such as ``What is the process to migrate from an Incubator to a Production cluster?'' requires an understanding of domain terminology that may be tacit or not readily available or usable for query classification. An important lesson learned is that while query classification can enhance the usefulness of a chatbot, achieving the necessary level of accuracy for such classification remains challenging.

\textbf{Design for scalability.} Many companies inevitably need to deploy chatbots internally due to privacy concerns. This makes scalability provisions a crucial consideration. While our chatbot was built primarily as a proof-of-concept for accuracy and was deployed on only one GPU for testing purposes, we ensured that our chatbot architecture supports horizontal scaling in a Kubernetes environment~\cite{selfadaptiveautoscaling2024}. If a chatbot is to become an adopted and widely used product, one must consider the possibility that several questions may need to be answered simultaneously. This requires multiple instances of the chatbot running in parallel, with each instance answering questions one at a time. We recommend that scalability needs be addressed early in chatbot design to identify bottlenecks and implement proper scalability mechanisms to support scaling up and down based on demand fluctuations.

%% file: conclusion.tex
\section{Conclusion}\label{sec:conclusion}
In this paper, we presented our experience developing a \comm{question-answering} chatbot for continuous integration and continuous delivery (CI/CD) at Ericsson. Through empirical evaluation using real-world CI/CD-related questions, we demonstrated that our chatbot provides useful answers to 87\% of queries, with over 60\% of the answers being fully correct. In future work, we plan to improve our chatbot's usability based on feedback from Ericsson. While our prototype shows feasibility in a production environment, improvements are paramount. To this end, we plan to conduct a user study to provide deeper insights into usability. Key areas for improvement include enhancing the chatbot's ability to handle complex logical queries and developing additional features, such as feedback loops, which are currently missing but are important for future success. \comm{In the longer term, we
would like to evolve our current chatbot from a question-answering tool into a smart agent capable of assisting with
the execution of CI/CD tasks based on user prompts.}